\documentclass[a4paper,runningheads]{llncs}
\usepackage[T1]{fontenc}
\usepackage{csquotes}
\usepackage{upquote}
\usepackage[left]{lineno}
\usepackage[UKenglish]{babel}
\usepackage{booktabs}
\usepackage{graphicx}
\usepackage{amsmath}
\usepackage{amsfonts}
\usepackage{paralist}
\usepackage{xcolor}
\usepackage{pdflscape}
\usepackage[style=lncs]{biblatex}
\usepackage{biblatex-readbbl}
\definecolor{lhWhite}        {RGB}{248,248,248}
\definecolor{lhBlack}        {RGB}{  8,  8,  8}
\definecolor{lhOrange}       {RGB}{243,107, 33}
\definecolor{lhGreen}        {RGB}{ 65,242, 34}
\definecolor{lhCyan}         {RGB}{ 34,169,242}
\definecolor{lhMagenta}      {RGB}{211, 34,242}
\definecolor{lhGold}         {RGB}{242,211, 34}
\definecolor{lhOrangeMedium} {RGB}{156, 69, 22}
\definecolor{lhGreenMedium}  {RGB}{42,156,22}
\definecolor{lhCyanMedium}   {RGB}{ 22,109,156}
\definecolor{lhMagentaMedium}{RGB}{135, 22,156}
\definecolor{lhGoldMedium}   {RGB}{156,135, 22}
\definecolor{lhOrangeDark}   {RGB}{ 71, 31, 10}
\definecolor{lhGreenDark}    {RGB}{ 19, 71, 10}
\definecolor{lhCyanDark}     {RGB}{ 10, 50, 71}
\definecolor{lhMagentaDark}  {RGB}{ 62, 10, 71}
\definecolor{lhGoldDark}     {RGB}{ 51, 86, 28}
\definecolor{lhLightGray}    {RGB}{204,204,204}
\definecolor{lhDarkGray}     {RGB}{ 68, 68, 68}
\usepackage{lh-listings}
\renewcommand{\lhdisplayfontsize}{\footnotesize}
\usepackage{lh-listings-isar}
\usepackage[autolanguage]{numprint}
\usepackage{units}
\usepackage{xspace}
\usepackage{hyperref}
\usepackage{subcaption}
\usepackage{longtable}
 \usepackage[LNCS,
  orcidicon,
  nobib,
  nourl,
  nocopyright,
  year=2023,
  publication={T.S. Hoang, {\'E}. Andr{\'e}, and L. Qin (Eds.): ICEFM 2026, LNCS},
  startpage={1},
]{authorarchive}

\usepackage{tikz}

\graphicspath{{figures/}{data/}}

\usepackage{hyperref}

\title{Interval-based Verification of Neural Networks: Formally Verified!\thanks{This work was supported by the Engineering and
Physical Sciences Research Council [grant number 670002170].}}
\authorrunning{A.D. Brucker and A. Stell}
\author{%
  \protect\href{http://brucker.ch/}{Achim
    D. Brucker\orcidID{0000-0002-6355-1200}} and 
  \protect\href{https://experts.exeter.ac.uk/37582-amy-stell}{Amy Stell\orcidID{0000-0003-0714-3269}}
}
\institute{%
  Department of Computer Science, 
  University of Exeter,
  Exeter, UK \texorpdfstring{\\}{}
  \email{{\{a.brucker,a.stell\}@exeter.ac.uk}
  }
}
\hypersetup{%
pdfauthor={Achim D. Brucker and Amy Stell},
pdftitle={Interval-based Verification of Neural Networks: Formally Verified! (Extended Version)},
pdfsubject={},
pdfkeywords={},
}

\begin{document}
\selectlanguage{UKenglish}%
\renewcommand{\bibname}{References}%
\renewcommand{\figurename}{Fig.}
\renewcommand{\abstractname}{Abstract.}
\renewcommand{\subsubsectionautorefname}{Sect.}
\renewcommand{\subsectionautorefname}{Sect.}
\renewcommand{\sectionautorefname}{Sect.}
\renewcommand{\figureautorefname}{Fig.}
\setlength{\parskip}{0pt}
\setlength{\parindent}{15pt}

\maketitle{}

\begin{abstract}
  Neural networks have demonstrated success in solving a wide range of problems, such as outperforming traditional image recognition approaches, thereby creating demand for their use in safety-critical systems. However, deployment in such systems requires formal verification to provide certification guarantees. As a consequence, we have seen a rise in formal verification techniques for neural networks; yet, these verification tools themselves typically lack formal correctness guarantees.

  In safety-critical applications, properties can often be expressed as interval constraints on network outputs, motivating interval-based verification approaches. One such approach has been successfully applied to the ACAS Xu aircraft collision avoidance benchmark. However, the correctness of this approach relies on informal mathematical arguments and a manually written, heavily optimised C implementation that has not been formally verified.

  In this paper, we present a mechanised verification methodology for neural networks in Isabelle/HOL. We prove that interval-based verification is sound by establishing three correctness properties: interval extension, inclusion isotonicity, and Lipschitz continuity. From our formal specification, we generate a trustworthy implementation in Haskell and demonstrate the verified methodology on ACAS Xu, producing machine-checked correctness proofs for safety properties.

\end{abstract}

\section{Introduction}\label{sec:introduction}\enlargethispage{\baselineskip}
\enlargethispage{\baselineskip}
Deploying neural networks in safety-critical systems requires formal certification: statistical testing provides no guarantees about behaviour on unseen inputs, making it insufficient for domains in which failures carry life-or-death consequences. This need has driven a substantial body of work on formal verification techniques for neural networks, and a number of automated tools now exist that can verify safety properties for networks of significant size~\cite{wang.ea:formal-security:2018,harapanahalli.ea:toolbox:2023}. Tools of this kind have featured prominently in the Verification of Neural Networks Competition (VNN-COMP)~\cite{kaulen:vnn-comp:2025}, which benchmarks their performance on safety-critical case studies including the ACAS Xu aircraft collision avoidance system~\cite{olson:acas-xu:2015}.

However, these tools are themselves unverified software artefacts: their correctness rests on informal mathematical arguments and heavily optimised, manually written C codebases~\cite{wang.ea:formal-security:2018}. For safety-critical certification, this is insufficient. A soundness bug in the verifier can produce a false guarantee of safety with no indication of failure. Formal methods practice demands that tools used in certification arguments be themselves subject to formal correctness guarantees.

Among the verification approaches in common use, interval-based methods are particularly well suited to safety-critical properties, which can frequently be expressed as bounds on network outputs~\cite{wang.ea:formal-security:2018,harapanahalli.ea:toolbox:2023}. These methods leverage interval arithmetic~\cite{moore.ea:introduction:2009} to compute sound over-approximations of network behaviour across continuous input regions. While the underlying mathematics of interval-based verification are already well established, existing implementations lack the mechanised formal guarantees required for true high-assurance certification.

In this paper, our focus is firmly on foundational trustworthiness rather than pushing the boundaries of benchmark scale. We address the gap between the formal guarantees demanded by certification and the informal foundations of existing tools by presenting a fully verified methodology for neural networks in Isabelle/HOL~\cite{nipkow.ea:isabelle:2002}. We formalise an interval-based approach inspired by~\cite{wang.ea:formal-security:2018}, building on our prior formalisations of neural networks~\cite{brucker.ea:network-verification:2023,brucker.ea:neural-networks-afp:2025} and interval arithmetic~\cite{brucker.ea:interval-arithmetic:2024,brucker.ea:interval-arithmetic-afp:2024}. We prove three key properties: interval extension, inclusion isotonicity, and Lipschitz continuity, establishing that interval-based verification is sound for any network whose activation functions satisfy these properties. From the formal specification, we generate a trustworthy Haskell implementation via Isabelle's code generator. We then demonstrate the complete pipeline on the ACAS Xu benchmark, producing machine-checked correctness proofs for safety properties as a proof-of-concept for our verified toolchain.

The novel contributions in this paper are: first, we formalise and verify a methodology, proving that interval-based verification is correct for neural networks with a large class of activation functions. Second, we develop proof engineering techniques that make this meta-verification more practical by creating Eisbach methods to automate the correctness proofs. Finally, we generate a trustworthy Haskell implementation directly from the formal specifications using Isabelle's code generator, demonstrating its applicability on the ACAS Xu case study to highlight the value of an unbroken chain of verified trust.

\section{Preliminaries}\label{sec:background} 
\enlargethispage{\baselineskip}
We will now briefly describe Isabelle/HOL (\autoref{sec:isabelle}), review the neural network formalisation from the Archive of Formal Proofs~\cite{brucker.ea:network-verification:2023,brucker.ea:neural-networks-afp:2025} (\autoref{sec:neural-networks}), and summarise the key concepts from our interval arithmetic formalisation~\cite{brucker.ea:interval-arithmetic:2024,brucker.ea:interval-arithmetic-afp:2024} (\autoref{sec:interval-arithmetic}).

\subsection{Isabelle/HOL}\label{sec:isabelle}
Isabelle~\cite{nipkow.ea:isabelle:2002} is a well-known interactive theorem
prover that has been used successfully in large-scale verification projects
(e.g.,~\cite{klein:operating:2009} presents the verification of an operating
system kernel using Isabelle/HOL). The formal language of Isabelle is
\emph{Higher-order logic} (HOL)~\cite{church:types:1940},
which is a classical logic based on a simple type system.
It provides the usual logical connectives
like \inlineisar|\<not> _|, \inlineisar|_ \<and> _|, \inlineisar|_ \<or> _|,
\inlineisar|_ \<longrightarrow> _| as well as the object-logical quantifiers
\inlineisar|\<forall> x. P x| and \inlineisar|\<exists> x. P x|. In contrast to
first-order logic, quantifiers may range over arbitrary types, including total
functions \inlineisar|f::\<alpha> \<Rightarrow> \<beta>| (where
\inlineisar|\<alpha>| and \inlineisar|\<beta>| are polymorphic type variables). 

Isabelle/HOL offers support for extending theories in a logically safe way: a
theory extension is \emph{conservative} if the extended theory is consistent,
provided that the original theory was consistent. Conservative extensions can
be \emph{constant definitions}, \emph{type definitions}, \emph{datatype
definitions}, \emph{primitive recursive definitions},
 and \emph{well-founded recursive definitions}.

\subsection{Neural Networks}\label{sec:neural-networks}
A feedforward neural network computes predictions by applying a sequence of transformations organised in layers. Each layer $\ell$ consists of a weight matrix $\omega_\ell$, a bias vector $\beta_\ell$, and an activation function $\varphi_\ell$. A layer computes $y_\ell = \varphi_\ell(\omega_\ell \cdot x_\ell + \beta_\ell)$ where $x_\ell$ is the input and $y_\ell$ is the output. The overall network prediction composes these layer transformations. For a network with $n$ layers: 
\begin{equation*}
  \text{predict}(x) = \varphi_n(\omega_n \cdot (\varphi_{n-1}(\omega_{n-1} \cdot \ldots \varphi_1(\omega_1 \cdot x + \beta_1) \ldots + \beta_{n-1}) + \beta_n))
\end{equation*}

We build on our previous formalisation of neural networks in Isabelle~\cite{brucker.ea:network-verification:2023,brucker.ea:neural-networks-afp:2025}, which represents networks as records containing a list of layers (e.g., input, output, dense, and activation) and a table of activation functions:
\begin{isar}
record ('a, 'b) neural_network_seq_layers = 
  layers ::  ('a, 'b) layer list
  activation_tab :: 'b \<Rightarrow> (('a list \<Rightarrow> 'a list) option)
\end{isar}

By using type variables \inlineisar{'a} and \inlineisar{'b}, we can reuse our formalisation of neural networks over reals (i.e., \inlineisar{'a} is instantiated with the type \inlineisar{real}). By instantiating \inlineisar{'a} with the type \inlineisar{real interval}, we obtain a representation over real intervals. Furthermore, \inlineisar{'b} is the type of activation functions.

\subsection{Interval Arithmetic and Lipschitz Continuity}\label{sec:interval-arithmetic}
Interval arithmetic computes with intervals $X = [\underline{x}, \overline{x}] \subseteq \mathbb{R}$ to over-approximate function behaviour across continuous input regions. For $f: \mathbb{R} \to \mathbb{R}$, an \emph{interval extension} $f_{I}: \mathcal{I}(\mathbb{R}) \to \mathcal{I}(\mathbb{R})$ must satisfy $x \in X \implies f(x) \in f_{I}(X)$. A function is \emph{inclusion isotonic} if it preserves interval containment: $X \subseteq Y \implies f_{I}(X) \subseteq f_{I}(Y)$. For inclusion isotonic functions, subdivision refines approximations: $f_{I}(X_1) \cup f_{I}(X_2) \subseteq f_{I}(X)$ where $X = X_1 \cup X_2$. A function $f$ is \emph{$L$-Lipschitz continuous} if $|f(x) - f(y)| \leq L \cdot |x - y|$. Lipschitz continuity bounds how much outputs can change relative to input changes.

Crucially, for Lipschitz continuous, inclusion isotonic functions, our mechanised proofs provide rigorous convergence guarantees: subdividing an interval into $n$ parts reduces the excess width (over-approximation) by a factor of $n$~\cite{brucker.ea:interval-arithmetic:2024}.

\section{Interval Properties of Neural Networks} \label{sec:interval-properties}
We establish three mathematical properties that enable sound, convergent verification.
\begin{inparaenum}
\item \emph{Interval Extension} (\autoref{sec:interval-extension-prop}): The interval network over-ap\-prox\-i\-mates the real-valued network, ensuring verification results are trustworthy.
\item \emph{Inclusion Isotonicity} (\autoref{sec:inclusion-isotonicity}): Dividing input intervals produces tighter bounds, enabling refinement-based verification.
\item \emph{Lipschitz Continuity} (\autoref{sec:lipschitz}): The network is bounded, guaranteeing convergence of refinement.
\end{inparaenum}

Our framework is parametric in the activation function: any activation function for which these three properties can be established is supported. In practice, this encompasses all Lipschitz continuous activation functions with a known interval extension. In \autoref{sec:verification} we discuss ReLU in the context of this example, however we have proven this property for a range of activation functions (e.g. ReLU, Identity, Softplus).
Combined with the convergence theorems from our interval arithmetic formalisation~\cite{brucker.ea:interval-arithmetic:2024}, these results provide formal guarantees that interval refinement will eventually converge to arbitrary accuracy for any neural network satisfying these properties.

\subsection{Formalising Interval Neural Networks} \label{sec:interval-networks}
We formalise interval neural networks in Isabelle/HOL by extending each component of a neural network in~\cite{brucker.ea:network-verification:2023} (e.g. activation functions, layers, and the prediction function) to operate on intervals.

\subsubsection{Interval Activation Functions.}\label{sec:interval-activation}

For an activation function $\varphi: \mathbb{R} \to \mathbb{R}$, we define the interval extension $\varphi_{I} : \mathcal{I}(\mathbb{R}) \to \mathcal{I}(\mathbb{R})$ by applying $\varphi$ to interval bounds. For instance, for the well-known ReLU activation function, we obtain:
\begin{equation*}
\text{relu}_{I}([\underline{x},\overline{x}]) = [\max(0,\underline{x}), \max(0,\overline{x})]
\end{equation*}

We generalise this construction for all activation functions:
\begin{isar}
definition activation_interval_of 
where
  activation_interval_of a = map_option (\<lambda> f. (\<lambda> l. map_interval_swap 
                        (zip (f (upper_l l)) (f (lower_l l))))) o a
\end{isar}
The function applies \inlineisar{f} to upper and lower bounds separately, zips the results, and uses \inlineisar|map_interval_swap| to swap bounds if necessary to ensure $\underline{x} \leq \overline{x}$; \inlineisar|map_option| handles missing activation function identifiers.

\newpage 

\subsubsection{Interval Layers.}\label{sec:interval-layers}
We now extend complete layers to operate on intervals:
\begin{isar}
fun layer_interval_of where
    \<open>layer_interval_of (In l) = In l\<close>
  | \<open>layer_interval_of (Out l) = Out l\<close>
  | \<open>layer_interval_of (Dense l) = Dense \<lparr>name = name l, 
       units = units l, \<phi> = \<phi> l, \<beta> = map interval_of (\<beta> l), 
       \<omega> = map (map interval_of)  (\<omega> l) \<rparr>\<close>
  | \<open>layer_interval_of (Activation l) = Activation l\<close>
\end{isar}
For dense layers, which perform weighted sums and bias addition, we convert each weight $\omega_{ij}$ and bias $\beta_i$ to a degenerate interval $[\omega_{ij}, \omega_{ij}]$ and $[\beta_i, \beta_i]$.
Input and output layers remain unchanged as they perform no computation. 
Activation layers keep their structure unchanged; the actual activation 
function will be looked up in the interval-extended activation table 
(defined in \autoref{sec:interval-activation}).

\subsubsection{Interval Network.}\label{sec:interval-network}
We construct interval neural networks as follows:
\begin{isar}
definition NN_interval_of 
where 
\<open>NN_interval_of N = (\<lparr>layers = map layer_interval_of (layers N), 
    activation_tab = activation_interval_of (activation_tab N)\<rparr>)\<close>
\end{isar}
This construction transforms every layer to its interval version (through \inlineisar|layer_interval_of|), and the activation function table to map the activation functions to their interval extensions (through \inlineisar|activation_interval_of|).

We also prove this construction preserves the structural properties of well-formed networks, enabling results proven for real-valued networks to transfer directly to their interval counterparts.

\subsubsection{Interval Prediction.}\label{sec:interval-prediction}

Given an input $x$, we convert $x$ to an interval vector by mapping each component to a degenerate interval, then apply the standard prediction function to the interval extended network:
\begin{isar}
definition 
  predict<bsub>seq_layer_int<esub> N xs = 
        predict<bsub>seq_layer_l<esub> (NN_interval_of N) (interval_of_list xs)
\end{isar}
where \inlineisar|interval_of_list| maps each component to $x_\ell \mapsto[x_\ell, x_\ell]$. Because the prediction function is polymorphic (type parameter \inlineisar|'a| in \autoref{sec:neural-networks}), it 
operates identically whether \inlineisar|'a| is instantiated to reals or 
intervals. As such, the function folds layer-wise prediction over the network's layers, applying interval arithmetic operations automatically.
For an input region $X$ (i.e., a list of intervals), we can compute the network's prediction over the entire input region. This allows us to verify properties for continuous regions of the input space.

\subsection{Interval Extension of the Prediction Function} \label{sec:interval-extension-prop}

Interval extension ensures that the interval-valued prediction function provides semantically equivalent results to the real-valued function. Formally, for any input $x$, the interval network's prediction on the degenerate interval $[x,x]$ must equal the original network's prediction: $\text{predict}_{I}(N, [x,x]) = \text{predict}(N, x)$, guaranteeing that interval approximations are conservative. We formally prove this property using structural induction over the layer types.

\subsubsection{Layer-wise Interval Extension.} \label{sec:layer-interval-ext}

We prove interval extension for each layer type. The proofs rely on interval extension properties of the underlying arithmetic operations and activation functions. Input and output layers perform identity transformations, trivially preserving interval extension: $[\text{id}(x), \text{id}(x)] = \text{id}([x,x])$. Activation layers apply an activation function element-wise. Interval extension holds provided the activation function itself is an interval extension:
\begin{isar}
lemma activation_layer_interval_extension:
assumes (Activation x1) \<in> set(layers N)
and     interval_extension_activation N 
shows (predict<bsub>seq_layer_l<esub>  (NN_interval_of N)
        (interval_of_list vs) (Activation x1)) =
      (interval_of_list(predict<bsub>seq_layer_l<esub> (N) vs (Activation x1)))
\end{isar}

The assumption \inlineisar|interval_extension_activation N| requires all 
activation functions in the network satisfy interval extension. For dense layers, which compute the weighted sum of the inputs (plus the bias), we establish:
\begin{isar}
lemma dense_layer_interval_extension:
assumes (Dense x1) \<in> set(layers N)
shows (predict<bsub>seq_layer_l<esub> (NN_interval_of N)
        (interval_of_list vs) (layer_interval_of (Dense x1))) =
      (interval_of_list(predict<bsub>seq_layer_l<esub> (N) vs (Dense x1)))
\end{isar}
The proof composes interval extension of matrix multiplication, bias addition, and activation to conclude the dense layer is an interval extension.

\subsubsection{Network-level Interval Extension.} \label{sec:network-interval-ext}

Having proven interval extension for individual layers, we establish it for complete networks. We formalise this within a locale~\cite{kammuller.ea:locales:1999} that captures the requirements:
\begin{isar}
locale 
interval_extension_neural_network_sequential_layers<bsub>l<esub> =
neural_network_sequential_layers<bsub>l<esub> +
constrains N :: ('a list, 'b, 'a list list) neural_network_seq_layers
assumes interval_extension: interval_extension_activation N
\end{isar}

This locale extends the locale (\inlineisar|neural_network_sequential_layers<bsub>l<esub>|) with one additional assumption: all activation functions in the network must be interval extensions (this is established by \inlineisar|activation_interval_of|, see~\autoref{sec:interval-activation}). 

\begin{isar}
lemma predict_interval_extension:
shows foldl (predict<bsub>seq_layer_l<esub> (NN_interval_of N))
  (interval_of_list vs) (layers (NN_interval_of N)) =
  interval_of_list (foldl (predict<bsub>seq_layer_l<esub> N) vs (layers N))
\end{isar}
The proof uses structural induction on the network: the base case handles networks with exactly two layers (input and output); the inductive case decomposes the network into its first layer and the remaining subnetwork, composing interval extension of each.

\subsection{Inclusion Isotonicity of the Prediction Function} \label{sec:inclusion-isotonicity}

Inclusion isotonicity ensures that the neural network prediction function 
preserves interval containment relationships. Formally, for input intervals 
$X \subseteq Y$:
\begin{equation*}
\text{predict}_{I}(N, X) \subseteq \text{predict}_{I}(N, Y)
\end{equation*}

\subsubsection{Layer-wise Inclusion Isotonicity.}
We establish inclusion isotonicity for each layer type. Input and output 
layers (being identity functions) trivially preserve the property. Activation layers preserve inclusion isotonicity provided that
the activation function is inclusion isotonic:\footnote{We formally prove inclusion isotonicity for a wide range of common activation functions, such as Identity, ReLU, Sigmoid, or SoftMax; for a full list, see~\cite{brucker.ea:neural-networks-afp:2025}.}
\begin{isar}
lemma activation_layer_inclusion_isotonicM:
assumes "inclusion_isotonicM (units x1) 
                 (the (activation_tab (NN_interval_of N) (\<phi> x1)))"
shows "inclusion_isotonicM (units x1) (\<lambda> vs . 
            (predict<bsub>layer_l<esub> (NN_interval_of N) vs (Activation x1)))"
\end{isar}

Dense layers compute weighted sums, add biases, and apply activation functions. Interval multiplication and addition are inclusion isotonic: if $X \subseteq Y$, then $\omega \cdot X \subseteq \omega \cdot Y$ and $X + \beta \subseteq Y + \beta$. Combined with inclusion isotonic activation functions, dense layers preserve the property:
\begin{isar}
lemma dense_layer_inclusion_isotonicM: 
assumes "inclusion_isotonicM (units x1) (the (activation_tab    
  (NN_interval_of N) (\<phi> x1)))"
and "length (map2 (+) (map (\<lambda>vs'. \<Sum>(x, y)\<leftarrow>vs'. x * y) 
  (map (zip xs) (\<omega> x1))) (\<beta> x1)) = units x1" 
shows "inclusion_isotonicM (in_deg_layer(Dense x1)) 
    (\<lambda> vs .  (predict<bsub>layer_l<esub> (NN_interval_of N) vs (Dense x1)))"
\end{isar}
The second assumption ensures dimensional consistency. 

\subsubsection{Network-level Inclusion Isotonicity.} \label{sec:network-inclusion}

Following the same compositional approach as for the network-level interval extension, we establish inclusion isotonicity for complete networks. We define a locale capturing the requirements:

\begin{isar}
locale 
inclusion_isotonic_neural_network_sequential_layers<bsub>l<esub> =
neural_network_sequential_layers<bsub>l<esub> +
constrains N:: ('a list, 'b, 'a list list) neural_network_seq_layers
assumes inclusion_isotonic: inclusion_isotonic_activation_tab<bsub>l<esub> 
  (NN_interval_of N)
and layers_consistent_interval: layers_consistent<bsub>l<esub> (NN_interval_of N) 
  (in_deg_NN (NN_interval_of N)) (layers(NN_interval_of N))
\end{isar}

The locale requires: (1) the network has consistent, well-formed layers, and (2) all activation functions in the activation table are inclusion isotonic. Within this locale, we prove:

\begin{isar}
lemma predict<bsub>seq_layer_l<esub>_inclusion_isotonicM:
assumes "layers_consistent<bsub>l<esub> (NN_interval_of N) 
  (in_deg_NN<bsub>l<esub>(NN_interval_of N)) (layers (NN_interval_of N))"
shows "inclusion_isotonicM (in_deg_NN<bsub>l<esub> (NN_interval_of N)) 
  (predict<bsub>seq_layer_l<esub> (NN_interval_of N))"
\end{isar}

Similarly to the interval extension proof, this uses structural induction, composing inclusion isotonic properties of individual layers.

\subsection{Lipschitz Continuity of the Prediction Function} \label{sec:lipschitz}

Lipschitz continuity quantifies how much network outputs can change relative to input changes. For the neural network prediction function $\text{predict}(N, \cdot)$ to be $L$-Lipschitz continuous:
\begin{equation*}
|\text{predict}(N, x) - \text{predict}(N, y)| \leq L \cdot |x - y|
\end{equation*}
This property is essential for interval refinement convergence. Combined with inclusion isotonicity, it guarantees that subdividing input intervals reduces over-approximation linearly with the subdivision width~\cite{brucker.ea:interval-arithmetic:2024}. Specifically, our previous work~\cite{brucker.ea:interval-arithmetic:2024} proves that for $L$-Lipschitz continuous, inclusion isotonic functions, the excess width (over-approximation error) after subdividing into $N$ parts is bounded by $2L \cdot w(X)/N$ where $w(X)$ is the interval width.

Establishing Lipschitz continuity for neural networks in Isabelle/HOL requires that our network representations instantiate the \inlineisar|real_normed_vector| type class, which provides the mathematical structure necessary to define norms and distances. For this reason, we prove Lipschitz continuity using the matrix-based sequential representation (which is semantically equivalent to the list-based model~\cite{brucker.ea:network-verification:2023}), which supports vector space operations.

\subsubsection{Layer-wise Lipschitz Continuity.}\label{sec:layer-lipschitz}
Common activation functions are Lipschitz continuous with small constants. For example, ReLU is 1-Lipschitz:
\begin{isar}
lemma relu_lipschitz:
  1-lipschitz_on (X::real set) (relu)
\end{isar}

This follows directly from ReLU's definition: $\text{relu}(x) = \max(0, x)$. For any $x, y$, we have $|\text{relu}(x) - \text{relu}(y)| \leq |x - y|$. Similarly, sigmoid and $\tanh$ are 1-Lipschitz (with maximum derivative 1), though we can compute tighter constants for specific intervals.

Input and output layers are 1-Lipschitz. For dense layers, we compose the Lipschitz properties of matrix multiplication, bias addition, and activation:

\begin{isar}
lemma dense_layer_lipschitz_on: assumes norm (\<omega> x1) \<leq> C 
and D-lipschitz_on ((\<lambda> y. y *fv \<omega> x1 + \<beta> x1) ` U) 
                (the (activation_tab N (\<phi> x1)))            
shows (C * D)-lipschitz_on U (\<lambda> vs. (predict_layer_f N vs (Dense x1)))
\end{isar}
The Lipschitz constant of a dense layer is the product of the weight matrix operator norm $C = \|\omega\|$ and the activation function's Lipschitz constant $D$, by the standard composition theorem for Lipschitz functions.

Activation layers inherit the Lipschitz constant of their activation function.

\subsubsection{Network-level Lipschitz Continuity.}

We establish Lipschitz continuity for complete networks by composing the layer-wise results. Our main theorem shows that if all layers (and activation functions) are Lipschitz continuous, then the entire prediction function is Lipschitz continuous.

The theorem demonstrates that sequential application of Lipschitz continuous layers preserves Lipschitz continuity, with the overall constant bounded by the product of individual layer constants. For a network with $n$ layers, where layer $\ell$ has weight matrix norm $|\omega_\ell|$ and activation function with Lipschitz constant $L_{\varphi_\ell}$, the overall network is $L$-Lipschitz continuous with:

\begin{equation*}
L \leq \prod_{\ell=1}^{n} (|\omega_\ell| \cdot L_{\varphi_\ell})
\end{equation*}
This result, combined with interval extension and inclusion isotonicity, provides the foundations for interval refinement with formal convergence guarantees.

\subsection{Correctness Guarantees}

We combine these properties into a verification framework. We define a locale that captures networks that are both inclusion isotonic and interval extension:

\begin{isar}
locale 
interval_extension_inclusion_isotonic_neural_network_sequential_layers<bsub>l<esub> 
= inclusion_isotonic_neural_network_sequential_layers<bsub>l<esub> 
+ interval_extension_neural_network_sequential_layers<bsub>l<esub>
\end{isar}
Within this combined locale, we establish:
\begin{isar}
theorem interval_predict_correct:
shows predict<bsub>seq_layer_l<esub> N vs \<in><bsub>I<esub> 
      predict<bsub>seq_layer_l<esub> (NN_interval_of N) (interval_of_list vs)
\end{isar}

This theorem states that the real-valued network's output on any specific input $v$ is contained within (\inlineisar|\<in><bsub>I<esub>|) the interval network's output on the corresponding degenerate interval $[v, v]$. Consequently, if we can prove a property $P$ holds for all values in the interval output, then $P$ must hold for the network's actual output on any input within the original interval region.

The proof combines our interval extension and inclusion isotonicity results, applying the convergence result from~\cite{brucker.ea:interval-arithmetic:2024} to the neural network context. This theorem provides the formal correctness guarantee of our verification approach.

\subsection{Interval Refinement}\label{sec:interval-refinement}
Whilst interval extension provides sound over-approximation, initial approximations may be too coarse for practical verification. Interval refinement addresses this: we subdivide the input space into smaller intervals, compute predictions on each subdivision, and take the union of results to obtain tighter bounds.

\subsubsection{Subdivision Strategy.}\label{sec:subdivision}

Given an input interval $X$ and a subdivision into $N$ parts $X_1, \ldots, X_N$ where $X = \bigcup_{\ell=1}^N X_\ell$, inclusion isotonicity guarantees:
\begin{equation*}
\bigcup_{\ell=1}^N \text{predict}_{I}(N, X_\ell) \subseteq \text{predict}_{I}(N, X)
\end{equation*}
Our framework guarantees soundness for \emph{any} valid subdivision strategy, provided the subdivisions cover the original interval without gaps.

This enables future optimisations of adaptive splitting strategies without requiring re-verification. The refinement provides tighter bounds, especially near discontinuities or regions of high non-linearity in the activation functions.

\subsubsection{Convergence Guarantee.}\label{sec:convergence}

By combining the three proven properties (interval extension, inclusion isotonicity, and Lipschitz continuity) with the convergence theorem from~\cite{brucker.ea:interval-arithmetic:2024}, we obtain formal convergence guarantees. 

For a neural network with Lipschitz constant $L$, subdividing an input interval $X$ of width $w(X)$ into $N$ equal parts produces an excess width (over-approximation error) bounded by:
\begin{equation*}
w(E) \leq 2L \cdot \frac{w(X)}{N}
\end{equation*}
This linear convergence ensures that sufficiently fine subdivisions will eventually produce arbitrarily tight approximations.

\subsubsection{Implementation and Workflow.}\label{sec:workflow}
The complete verification pipeline proceeds in five stages, entirely within or orchestrated from Isabelle/HOL:
\begin{compactenum}
  \item \textbf{Network import.} Each neural network is imported from TensorFlow JSON format using our existing import mechanism~\cite{brucker.ea:network-verification:2023}, which automatically establishes locale membership and structural well-formedness within Isabelle.
  \item \textbf{Property discharge.} Interval extension, inclusion isotonicity, and Lipschitz continuity are established for the imported network. Because all required activation function properties have been pre-proved, this step is fully automated via Eisbach proof methods (see \autoref{sec:eisbach}).
  \item \textbf{Code export and execution.} Isabelle's \inlineisar|export_code| directive generates a Haskell programme from the formal specification. This programme is executed externally to compute the interval partition for each verification goal, performing iterative uniform subdivision up to a configurable depth.
  \item \textbf{Interval import.} The computed partition is loaded into Isabelle as a formal HOL definition using the \inlineisar|import_intervals| command (see \autoref{sec:import-intervals}), closing the pipeline within the trusted kernel.
  \item \textbf{Verification.} Isabelle verifies two machine-checked lemmas per goal: that the imported partition is a valid cover of the original input region, and that the safety property holds over all sub-intervals. Both proofs proceed by \inlineisar|eval|.
\end{compactenum}
The correctness of the overall verification result does not depend on the correctness of the generated Haskell code: step~5 re-establishes partition validity from scratch within Isabelle, so even an incorrect or optimised external computation cannot introduce unsoundness. The full pipeline is detailed in \autoref{sec:verification}.

\section{Verification Methodology and Evaluation}\label{sec:verification}

This section presents the proof engineering infrastructure that makes our verified methodology practical to apply, and demonstrates its instantiation on the ACAS Xu benchmark~\cite{olson:acas-xu:2015}, a standard benchmark for neural network verification tools featured in VNN-COMP since 2020. ACAS Xu represents a safety-critical application where neural networks make collision avoidance decisions for unmanned aircraft, making formal verification essential for certification.

\subsection{Eisbach Proof Methods}\label{sec:eisbach}

Isabelle's Eisbach framework allows proof methods to be defined from existing tactics~\cite{matichuk:eisbach:2016}. We define reusable Eisbach methods that automate the interval extension and inclusion isotonicity proofs for all 45 ACAS Xu networks, parametrised by the network-specific constants produced during import. Each network's proof reduces to a single method invocation; the same pattern is applied uniformly across all 45 networks for both properties, making the proof engineering cost constant per network regardless of scale.

\subsection{Code Generation} \label{sec:code-gen}
\enlargethispage{\baselineskip}
The interval partition is computed using Haskell code generated from the Isabelle specification. For each network \inlineisar|Network<bsub>i,j<esub>| and property $\varphi_k$ we compute:

\begin{isar}
definition "Network_i_j_Property_k_Generate_Intervals =
  process_property 10000 Network_i_j.NeuralNet check_pk
                   split_interval_lists Input_k"
\end{isar}

where \inlineisar|process_property| performs iterative uniform subdivision up to a maximum depth (10000 in this case), stopping early when the property predicate \inlineisar|check_pk| is satisfied over all sub-intervals. Code for the computation 
can be run either within Isabelle or as a stand-alone programme. 

It is important to note that the correctness of our verification does not rely on the correctness of this generated code, as for each computed split, Isabelle can easily verify its correctness: we only need to formally prove that the list of interval splits covers the original interval, using the predicate:
\begin{isar}
definition check_intervals :: "'a interval list list 
                              \<Rightarrow> 'a interval list list \<Rightarrow> bool" where
"check_intervals splits original 
 = ([concat (map mk_mInterval_non_ovl (transpose splits))] = original)"
\end{isar}

Consequently, we can exploit even unsafe optimisations of Isabelle's code generator to improve the performance of the interval split computation, without affecting soundness: Isabelle validates the correctness of any computed splitting via \inlineisar|check_intervals|. For the ACAS Xu application we export Haskell code and configure the code generator to map mathematical reals to IEEE754 double-precision floats. Although this mapping is unsound in general, it does not affect the verification results since correctness of the splitting is re-checked in Isabelle.

\subsection{Interval Import}\label{sec:import-intervals}

The computed partition must be loaded back into Isabelle and registered as a formal HOL definition, closing the pipeline entirely within Isabelle. We implement this as a new Isabelle top-level command:

\begin{isar}
import_intervals "path/to/network_1_2_Property_1_Intervals.txt"
  defining network_1_2_Property_1_intervals
\end{isar}

The file contains one line per input, each encoding a list of real 
intervals in the format \texttt{[(lo\textsubscript{1},hi\textsubscript{1}),(lo\textsubscript{2},hi\textsubscript{2}),\ldots]}, where bounds are decimal numbers in scientific notation as printed by Haskell's \texttt{hPrint}. The command parses the file, produces a HOL definition of type \inlineisar|real interval list list|, and makes it immediately available for use in proofs.

\subsection{ACAS Xu Benchmark}
\enlargethispage{\baselineskip}
\subsubsection{Network Architecture.}\label{sec:network-arch}

Each ACAS Xu network is a fully connected feedforward network with 6 hidden layers of 50 neurons each, using ReLU activations. The networks take five continuous inputs encoding the relative geometry and velocities of two aircraft (separation distance, heading angles, and speeds).

Two additional state variables, $\tau$ (time until loss of vertical separation) and $a_{\text{prev}}$ (previous advisory), determine which of the 45 networks to use. Networks are named $N_{i,j}$ where $i \in \{1, \ldots, 5\}$ indexes the previous advisory $a_{\text{prev}} \in \{\text{COC}, \text{WL}, \text{WR}, \text{SL}, \text{SR}\}$ and $j \in \{1, \ldots, 9\}$ indexes the time until vertical separation $\tau \in \{0, 1, 5, 10, 20, 40, 60, 80, 100\}$ seconds.

Each network outputs five scores corresponding to turning advisories: COC (clear of conflict), WL (weak left, 1.5 deg/s), WR (weak right, 1.5 deg/s), SL (strong left, 3.0 deg/s), and SR (strong right, 3.0 deg/s). The advisory with the minimum score is selected as the recommended action.

\subsubsection{Network Import and Property Formalisation.}\label{sec:property-formalisation}

We import all 45 ACAS Xu networks using our TensorFlow import mechanism~\cite{brucker.ea:network-verification:2023,brucker.ea:neural-networks-afp:2025}:
\begin{isar}
import_TensorFlow Network_1_1 
  file ./Networks/Network_1_1/model/model.json as seq_layer_list
\end{isar}

The import automatically establishes that each network satisfies the requirements of the \inlineisar|neural_network_sequential_layers| locale, verifying structural correctness including layer compatibility and activation function validity.

The verification properties are formalised as Boolean predicates over interval outputs. For $\varphi_1$ (the COC advisory score remains below 1500 when the intruder is distant and significantly slower), we verify that the upper bound of the COC score interval does not exceed the threshold:

\begin{isar}
definition check_p1 output = (upper (output ! 0) \<le> 1500)
\end{isar}

For $\varphi_2$ (COC is never maximal under the same conditions), we ensure COC is not maximal by requiring some other advisory has a lower bound exceeding COC's upper bound:

\begin{isar}
definition check_p2 output =
(\<exists>i < length output. i \<noteq> 0 \<and> upper (output ! 0) < lower (output ! i))
\end{isar}

Properties $\varphi_3$ and $\varphi_4$ (COC is not minimal when the intruder is directly ahead), both verify COC is not minimal, requiring some advisory has an upper bound below COC's lower bound:

\begin{isar}
definition check_p3_p4 output =
(\<exists>i < length output. i \<noteq> 0 \<and> upper (output ! i) < lower (output ! 0))
\end{isar}

These predicates are executable, meaning that Isabelle can evaluate them automatically on computed intervals.

\subsection{Verification Results}\label{sec:results}
\enlargethispage{\baselineskip}

The primary contribution of this work is a \emph{formally verified methodology}: the correctness of our interval-based verification approach is guaranteed by machine-checked proofs in Isabelle/HOL, independently of how efficiently any particular instance runs.
The ACAS Xu case study serves as a proof-of-concept demonstrating that the methodology is practically applicable, not a claim of competitive performance against state-of-the-art solvers.

For each verified goal, we produce two machine-checked lemmas. The first establishes that the computed interval partition is valid, i.e., the subdivisions are contiguous and cover the original input space exactly:

\begin{isar}
lemma check_Network_i_j_Property_k:
shows check_intervals Network_i_j_Property_k_Gen_Intervals Input_k
by eval
\end{isar}

The second lemma proves the safety property holds over all computed intervals:
\begin{isar}
lemma prove_Network_i_j_Property_k:
shows check_pk (predict<bsub>seq_layer_i<esub> (NN_interval_of Network_i_j.NeuralNet) 
  Network_i_j_Property_k_Generate_Intervals)
by eval
\end{isar}

Together, these two lemmas constitute a complete correctness proof for the safety property: the partition is formally verified to be valid, and the property is formally verified to hold over it. 

Furthermore, this modular approach enables parallel verification of multiple properties using Isabelle's session management. Out of the 86 network-property combinations for which we generate interval partitions (across properties $\varphi_1$--$\varphi_4$ of the ACAS Xu benchmark), we successfully verify 63; since each verified combination yields the two lemmas described above, this amounts to 126 machine-checked proofs in total. Of these, 104 proofs (52 combinations) complete within 75 minutes of total processing time; the remaining 22 proofs (11 combinations) take up to approximately 9 hours.
While further investigations are required, the main performance limitation appears to be the memory inefficiency of the generated Haskell code, which is primarily influenced by the number of interval splits. We identify significant optimisation potential here for future work, for example a more optimised splitting strategy; crucially, because Isabelle independently re-verifies the interval geometry, developers can aggressively optimise this external computation without risking the soundness of the overall verification pipeline.

While these empirical results are impressive for a fully mechanised approach, we reiterate that the primary contribution of this work is the verified methodology itself, i.e., the machine-checked correctness of the interval-based verification approach, which remains independent of the computational strategy used to produce the partition. To demonstrate broader applicability, the framework has also been applied successfully to a further case study beyond ACAS Xu (with non-linear activation functions~\cite{stell:phd:2025}).

\section{Discussion}\label{sec:discussion}

A key strength of our framework is the strict separation between the verified core methodology and the computational refinement strategy. Because our mathematical guarantees (interval extension, inclusion isotonicity, and Lipschitz continuity) are independent of how the input space is subdivided, rather than of the underlying methodology. Several targeted refinement strategies could address it without touching the trusted core, such as: activation-phase-guided splitting, which avoids mixing ReLU phases within a sub-interval and substantially tightens the interval extension in piecewise-linear regions; gradient- or sensitivity-guided adaptive subdivision, which concentrates effort on regions of high local variation instead of splitting uniformly; and branch-and-bound-style refinement, which stops subdividing a sub-region as soon as it is established safe, rather than subdividing the whole domain to a fixed depth. The only requirement is that the new subdivision covers the original input space without gaps, a structural property trivially established via our \inlineisar|check_intervals| predicate in Isabelle/HOL.

Like all over-approximation approaches, our methodology is \emph{sound but not complete}, trading completeness for the rigorous formal certifiability required in safety-critical applications. Scalability is practically constrained by the size of the interval partition passed to Isabelle; because the required subdivision precision (dictated by the Lipschitz constant) grows multiplicatively with network depth, implementing the more targeted external splitting strategies mentioned above would directly improve performance.

Our formalisation targets feedforward networks built from dense and activation layers; extending it to convolutional, recurrent, or attention-based architectures would require formalising the corresponding layer types and re-establishing interval extension, inclusion isotonicity, and Lipschitz continuity for them. While our proofs establish these properties for a range of activation functions beyond ReLU (\autoref{sec:inclusion-isotonicity}), our end-to-end evaluation exercises only ReLU networks, since all 45 ACAS Xu networks use exclusively ReLU activations.

Finally, an ``elephant in the room'' affecting most neural network verification is the use of IEEE754 floating-point numbers, which are \emph{not} faithful, bounded representations of mathematical reals. We verify correctness at the level of mathematical reals; bridging the gap to floating-point execution semantics is non-trivial future work. While basic real arithmetic is inclusion isotonic, IEEE754 operations are not, and they even lack universal algebraic properties such as associativity. Addressing this requires bit-precise verification techniques~\cite{yu:floating-point-afp:2013,boldo:hal-01632617}, a challenge further complicated by undocumented hardware-level optimisations in GPUs that might swap binary arguments. Consequently, bridging the gap between mathematical reals and hardware execution remains an essential area of future research for all AI verification methodologies.

\section{Related Work}\label{sec:related-work}

The closest related work is the interval-based verification approach for neural networks presented by Wang et al.~\cite{wang.ea:formal-security:2018}. In contrast to our approach, which is based on mechanised proofs and verified generated code, they rely on paper-and-pencil proofs and a heavily optimised C implementation. Other interval-based approaches have been presented by Harapanahalli et al.~\cite{harapanahalli.ea:toolbox:2023}. Crucially, these approaches share a common limitation: the absence of formal guarantees regarding the correctness of the verification methodology itself.

Several recent works have addressed the problem of \emph{verifying the verifier}, a category closely related to our contribution. Tobler et al.~\cite{tobler.ea:cav:2025} present a formally verified robustness certifier for neural networks in Dafny, also based on interval arithmetic; their approach shares our emphasis on a formally correct methodology, though their proof infrastructure and target language differ. Their current implementation, however, targets a narrower class of networks: dense feedforward ReLU networks without biases. Our formalisation and evaluation, by contrast, handle biased dense layers and arbitrary activation functions. ProveSound~\cite{singh.ea:provesound:2025} provides a procedure that automates the soundness verification of abstract interpretation-based DNN certifiers, covering interval analysis as a special case; unlike our work, it operates by checking certifiers against a symbolic DNN model rather than verifying the analysis engine within an ITP. Desmartin et al.~\cite{desmartin.ea:itp:2025} present a certified proof checker for Marabou in Imandra, an industrial ITP distinct from Isabelle/HOL, targeting SAT-based refutations rather than over-approximation proofs. Duong et al.~\cite{duong.ea:dnn-proofs:2025} introduce proof-producing Branch-and-Bound verification, in which the verifier emits a formal proof object that is checked by a separate minimal checker; this approach applies to ACAS Xu and is complementary to ours in addressing the trustworthiness gap in automated DNN verification. Our approach is distinct from all of these: rather than building a checker for an external tool's output, we verify the \emph{methodology}; specifically, the three key mathematical properties and their composition, directly in Isabelle/HOL, and generate a trusted executable implementation from this formal specification. 

Overall, interactive theorem provers (ITPs) remain underexplored for neural network verification, despite offering the highest assurance levels. Bentkamp et al.~\cite{bentkamp:deep:2016} formalise deep learning expressiveness in Isabelle/HOL. Aleksandrov and V\"{o}llinger~\cite{aleksandrov.ea:formalising:2023} formalise piecewise affine activation functions in Coq/Rocq, implementing and proving properties of ReLU. Finally, the Vehicle framework~\cite{daggitt.ea:vehicle:2025} bridges the gap between automated verifiers and ITPs through a domain-specific language that compiles to both targets, addressing the embedding gap that arises when combining neural and symbolic proof components.

\section{Conclusion}\label{sec:conclusion}
\enlargethispage{\baselineskip}
We have presented a formalisation of the interval-based verification approach for neural networks in Isabelle/HOL. The automation support that we provide turns this formalisation into a comprehensive verification framework developed within an interactive theorem prover. Our contribution establishes that neural networks possess the mathematical properties necessary for sound interval-based verification by proving three key results: interval extension, inclusion isotonicity, and Lipschitz continuity for neural network prediction functions.

These properties, proven compositionally from activation functions through individual layers to complete networks, enable the application of convergence theorems from interval arithmetic to reduce over-approximation errors~\cite{brucker.ea:interval-arithmetic:2024}. The result is a verification framework with formal guarantees about both the verification results and the verification process itself.

The separation between verified methodology and computational strategy distinguishes our approach: improvements to efficiency can be implemented and evaluated independently without compromising the formal guarantees that make the framework suitable for certification of safety-critical systems.

\subsubsection*{Availability.} 
The formalisation and case study for Isabelle 2025-2, as described in this
paper, are available to view on
Zenodo~\cite{brucker.ea:zenodo-interval-nn-verification:2026}. A maintained version that also will receive updates (including being ported
to future versions of Isabelle) is currently under review at the Archive of Formal Proofs
(AFP)~\cite{brucker.ea:afp-interval-nn-verification:2026} and will be published using a BSD-license
(SPDX-License-Identifier: BSD-3-Clause).

\printbibliography

\typeout{get arXiv to do 4 passes: Label(s) may have changed. Rerun}
\end{document}